\documentclass[conference]{IEEEtran}
\IEEEoverridecommandlockouts
\usepackage{cite}
\usepackage{amsmath,amssymb,amsfonts}
\usepackage{algorithmic}
\usepackage{graphicx}
\usepackage{soul}
\usepackage{textcomp}
\usepackage{xcolor}
\def\BibTeX{{\rm B\kern-.05em{\sc i\kern-.025em b}\kern-.08em
    T\kern-.1667em\lower.7ex\hbox{E}\kern-.125emX}}
\begin{document}

\title{Simplified Stability Assessment of Power Systems with Variable-Delay Wide-Area Damping Control
}

\author{\IEEEauthorblockN{K. K. Gajjar, Kaustav Dey and A. M. Kulkarni}
\IEEEauthorblockA{\textit{Department of Electrical Engineering}, \textit{Indian Institute of Technology Bombay, Mumbai, India} \\
kkgajjar@iitb.ac.in, kaustavd@iitb.ac.in, anil@ee.iitb.ac.in}
}

\maketitle

\begin{abstract}
Power electronic devices such as HVDC and FACTS can be used to improve the damping of poorly damped inter-area modes in large power systems. This involves the use of wide-area feedback signals, which are transmitted via communication networks. The performance of the closed-loop system is strongly influenced by the delay associated with wide-area signals. The random nature of this delay introduces a switched linear system model. \textcolor{blue}{The stability assessment of such a system requires linear matrix inequality based approaches. This makes the stability analysis more complicated as the system size increases. To address this challenge, this paper proposes a delay-processing strategy that simplifies the modelling and analysis in discrete-domain. In contrast to the existing stability assessment techniques, the proposed approach is advantageous because the stability, as well as damping performance, can be accurately predicted by a simplified analysis.} The proposed methodology is verified with a case study on the 2-area 4-machine power system with a series compensated tie-line. The results are found to be in accordance with the predictions of the proposed simplified analysis.


\end{abstract}


\section{Introduction}
 In large grids,  poorly damped low-frequency modes are frequently observed over a wide-area and require  special attention. These low-frequency modes can be damped by  power electronics actuators such as HVDC systems and FACTS. Signals synthesized from wide-area measurements may be used to selectively damp critical modes \cite{vedanta}. 


Wide-area signals, however, suffer from delays which typically lie between a few milliseconds to a few hundreds of milliseconds. These delays can affect the performance of the damping controllers and may also destabilize the system~\cite{delay_stab}. The delay involved with Wide-Area Measurement Systems~(WAMS) is variable and depends on different factors such as type of PMU, communication network and the  protocol used for data transmission~\cite{d_2015}. 
\par Previous research mainly focused on three different aspects: (i) Stability analysis of existing wide-area controller under various delay conditions. The system stability analysis considering fixed time delay is presented in~\cite{fd2_2008,wilches2017effect}. To assess the stability under variable delay, a Lyapunov-based method was proposed in \cite{d_2014}, which  estimated the delay margin. The delay margin is defined as the maximum time delay up to which the system remains stable. (ii) Development of a strategy to compensate for the adverse effects of time-delay~\cite{delay_em1}. (iii) A robust Wide Area Damping Controller~(WADC) design considering the delay associated with WAMS~\cite{d_2014}. 

While assessment of stability under fixed or time-varying delay has been considered earlier, as discussed, analytical  characterization of the damping performance in terms of bounds  on the performance has not been addressed earlier.  \textcolor{blue}{Motivated by this, the main contributions of this paper are (i) to propose a strategy to handle random variable delay, data packet dropout, and data packet disordering in order to make the behaviour more predictable, (ii) simplified analysis to assess the stability and effective damping using this strategy through characterization of the bounds on the system behaviour.} The stability aspects are compared with the results of the linear matrix inequality (LMI) based approach.

\section{Proposed Strategy for variable time delay}
\begin{figure}[!h]
    \centering
    \includegraphics[scale=0.23]{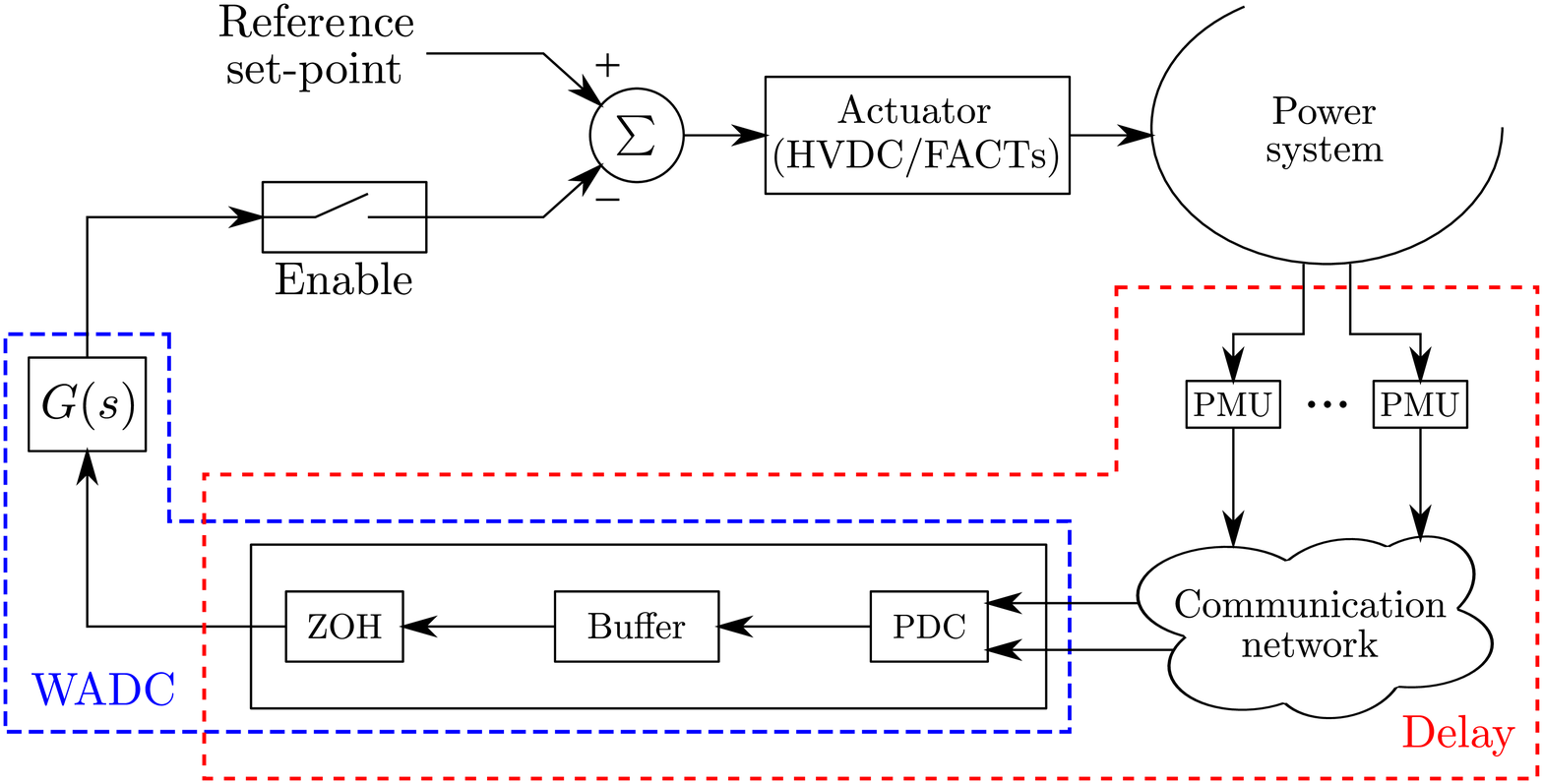}
    \caption{Schematic of a power system with a WADC}
    \label{Fig:cl}
\end{figure}

The schematic of a power system with a WADC is shown in Fig.~\ref{Fig:cl}. The measurements at different locations are captured and time-stamped by the respective PMUs and are then transmitted to the remotely located WADC via a communication network. The WADC modulates the actuator set-points~(e.g.~power reference in an HVDC converter) based on the PMU measurements in order to damp the controllable inter-area modes. The transmission of PMU measurements over the communication channel will introduce a variable time-delay in the feedback signals, which will affect the stability of the closed-loop system. Therefore, the delay needs to be modelled appropriately in order to assess the system stability. 

\par The total delay ($\tau_d$) associated with the measurements can be expressed as $\tau_d = \tau_p + \tau_c + \tau_o$ where (i) $\tau_p$ is the delay associated with the internal algorithm of the PMU. $\tau_p$ is usually random, and its variation can be modelled
as a normal distribution~\cite{d_2015}, (ii) $\tau_c$ 
is the delay due to the communication network latency. It depends on the network framework, and (iii) $\tau_o$ is the operational delay which is introduced by the digital processors in the WADC.

\par In addition, data packet dropout~\cite{data_dropout} and data packet disordering~\cite{data_disordering} is also possible. The variable nature of the delay makes the modelling complicated. The modelling in the discrete-time domain can be simplified if specific strategies are adopted to handle the variable time delay. The details of this aspect are presented in the subsequent sections.  

\subsection{Delay Modelling}
This paper proposes a strategy for handling the wide-area measurements, which can have variable delay. This strategy first converts the continuous-time (CT) variable delay into discrete-time (DT) variable delay and then uses ZOH to obtain a DT system with fixed sampling interval. 
 In practice, the PMU reporting rate is fixed (say $h$). Therefore, the output of ZOH is allowed to change at an interval of $h$. The controller may require measurements from multiple locations in order to generate the feedback signal. Although the delay in each of these channels can be different, the proposed strategy converts these signals into a composite signal of a particular delay. The strategy is divided into two steps.
\subsubsection{Time alignment and buffer storage}
In this step, Phasor Data Concentrator~(PDC) will continuously accept the incoming PMU data packets, then extract the time-stamp information and store the data in the buffer. In case of data packet disorder, only the latest sample is stored.
\subsubsection{Data processing and Zero Order Hold (ZOH)}
In this step, the composite signal is then generated using the buffered data. The {\em complete} available latest set of synchronous samples are used to generate the composite signal. The ZOH block then holds the signal for one time-step ($h$). 
\par
\begin{figure}[!h]
    \centering
    \includegraphics[scale=0.8]{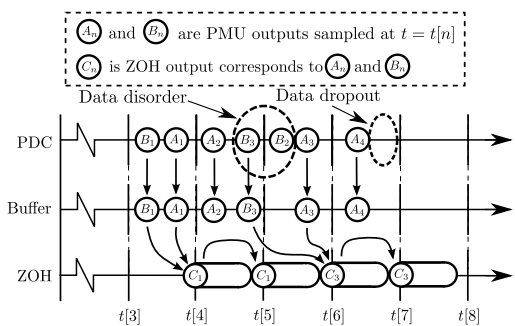}
    \caption{Illustrative example of the proposed strategy}
    \label{Fig:str}
\end{figure}

 As an illustration, Fig. \ref{Fig:str} provides the working example of the  proposed strategy. Consider two PMUs are transmitting data samples $A_n$ and $B_n$ respectively, which are time-stamped at $t = t[n]$. Let $C_n$ denote the ZOH output corresponding to $A_n$ and $B_n$. The nominal data generation and transmission latency is assumed to be $2h$ i.e. the data sampled at $t[n]$ are usually available to the PDC at or after $t[n+2]$.
\par \underline{Normal operation:} At time $t[1]$, the PMU samples $A_1$ and $B_1$ reach the PDC sometime between $t[3]$ and $t[4]$. Since a new output is to be generated only at the sampling instants, the measurements are stored in a buffer. The signal $C_1$ is then generated from $A_1$ and $B_1$ at $t[4]$. The output of the ZOH is passed to the damping controller at $t[4]$. 
\par \underline{Data packet disorder:} Consider a case wherein $A_2$ is already available in the buffer at $t[5]$. However, due to data packet disordering, $B_3$ arrives before $B_2$, and therefore, $B_2$ is discarded. The ZOH output $C_1$ is held on to till $t[6]$ until the corresponding sample of the other PMU~($A_3$) is received.
\par \underline{Data packet dropout:} Consider a case wherein the sample $B_4$ does not reach the PDC even after $t[6]$. In such a case, the output of the ZOH is held on to its previous output $C_3$. Therefore, data packet dropout and data packet disordering scenarios can be modelled as signals with larger delay.

\par In the proposed strategy, the output of the ZOH changes only at the sampling instants~(uniform sampling). 
The variable delay of the wide-area system can, therefore, be conveniently represented in the DT domain with fixed sampling intervals. If $u_d[K]$ is the input measurement signal and $y_d[K]$ is the delayed output sample at $t[K]$, then a $n$-step delay can be represented by $z^{-n}$ in the DT $z$-domain. An equivalent state-space model of the delay can be given as follows.
\begin{equation*} 
       \begin{aligned}
       x_d[K+1] = A_d x_d[K]+B_d u_d[K], \, y_d[K]  = C_d x_d[K]
       \end{aligned}
\end{equation*}
\textcolor{red}{Note that the order of the delay model~(dimension of $x_d$) is dependent on the maximum permissible delay~(as multiple of discretization time-steps)}. This can be chosen to be the fixed delay beyond which the system becomes unstable. As an example, for maximum delay of 3$h$, the state-space matrices are as follows.
\[
A_d=
\begin{bmatrix}
0 & 1 & 0\\
0 & 0 & 1\\
0 & 0 & 0
\end{bmatrix}
,~B_d = 
\begin{bmatrix}
0\\0\\1
\end{bmatrix}
,~{C_d}^T =
\begin{bmatrix}
\alpha_3 \\ \alpha_2 \\ \alpha_1\\
\end{bmatrix}
\]
Note that each $\alpha_i$ is either 1 or 0, with $ \sum_{i=1}^3{\alpha_i}=1$. In this state-space realization, $A_d$ and $B_d$ are fixed while $C_d$ depends on the delay. For example, $\alpha_1 = 1$ corresponds to the case of minimum delay~($h$) and $\alpha_3 = 1$ corresponds to maximum delay~($3h$) scenario. \textcolor{blue}{An intuitive understanding of this realization is as follows: the structures of $A_d$ and $B_d$ emulate that of a set of shift registers, and $C_d$ selects an element from that set of registers.} In order to interface the rest of the system with this delay model, the DT modelling of the power system is also required, which is presented next.

\subsection{Power system modeling}
Consider the linearised state-space model of the power system as given below:
$$ \Dot{x}(t) = Ax(t)+Bu(t), \,\, y(t) = Cx(t)+Du(t) $$

For large systems, there are several swing modes. \textcolor{red}{In order to analyze a particular mode of interest, different model order reduction techniques can be used to reduce the system size~\cite{kevin}. Therefore, the reduced order model of the power system can be considered to represent only the poorly-damped inter-area mode.}

The CT system is discretized using the Trapezoidal rule. \textcolor{blue}{Since the delay modelling is done at a time-step of $h$~(PMU reporting rate), the discretization of the power system model is also done with the same time-step.} The DT state-space model of the power system can be represented as follows.
\[
       \begin{aligned}
       x[K+1] & = A_p x[K]+B_{p}( u[K]+ u[K+1])\\
       y[K] & = C x_p[K] + D u[K]
       \end{aligned}
\]
where $x[K] = x(Kh)$. The DT state-space matrices are related to the CT state-space matrices as follows. 
\begin{align} \label{Eq:DT_CT_A_matrix}
    A_p = \left( 2I-Ah\right)^{-1} \left(2I+Ah \right), B_{p} = \left( 2I-Ah \right)^{-1} Bh
\end{align}
The relationship between the DT eigenvalues ($\mu_i$) and the corresponding CT eigenvalues ($\lambda_i$) are derived from~\eqref{Eq:DT_CT_A_matrix} and are given as follows.
\begin{align} \label{Eq:trapezoidal_rule}
    \mu_i = (2+ \lambda_i h)/(2-\lambda_i h) \approx e^{\lambda_i h} \quad \text{ if $|\lambda_i h| \ll 2$}
\end{align}
The frequency of the inter-area modes are usually such that $\lambda_i h \ll 2$. The following points are important:\\
(i) If the CT system is stable, i.e. real part of $\lambda_i < 0$, then $|\mu_i| < 1$, i.e. the DT system will also be stable.\\
(ii) The damping of the CT eigenvalue~(real part of $\lambda_i$) is reflected in the magnitude of $\mu_i$. \\



\subsection{Closed-loop system model}


The DT state-space model of the delay and the power system are interfaced~(as shown in Fig.~\ref{Fig:cl}), and the closed-loop system is obtained. Let $A_c$ denote the closed-loop DT state-transition matrix. For $N$ different delays, there are $N$ different $C_d$ matrices, resulting in $N$ different $A_c$ matrices. The corresponding $A_c$ matrices are denoted by $A_{C1},~A_{C2},...,~A_{CN}$ respectively. This forms a switched DT linear time-invariant (LTI) system with $A_{Ci}$ as its  ``switching states''.


\begin{figure}[!h]
    \centering
    \includegraphics[scale=0.35]{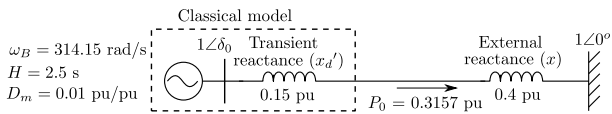}
    \caption{Single line diagram of a SMIB system}
    \label{Fig:SMIB_ex}
\end{figure}
\par \underline{Illustrative Example:} Consider the  single-machine infinite bus~(SMIB) system shown in Fig.~\ref{Fig:SMIB_ex}. The generator is modelled by the classical model. The generator data, network parameters, and the equilibrium condition is indicated in the same figure. The CT state-space matrices are as follows. 
\[
\begin{bmatrix}
\Delta \dot{\delta} \\ \Delta \dot{\omega}
\end{bmatrix}=
\begin{bmatrix}
0 & 1\\
-112.5 & -0.628
\end{bmatrix} \begin{bmatrix}
\Delta \delta \\ \Delta \omega
\end{bmatrix}
+
\begin{bmatrix}
0\\-62.83
\end{bmatrix} \Delta T
\]
where $T$ is the electro-magnetic torque output of the generator. The CT eigenvalue of the system is $\lambda = -0.314 \pm j 10.6$. Consider that the generator has an auxiliary controller which uses the speed~($\Delta \omega$) measurements at a reporting rate of $50$~Hz. Note that $|\lambda h| \ll 2$.  The system is discretized with $h = 0.02$~s~(corresponding to 50~Hz) to obtain the following DT state-space matrices. 
\[
A_p=
\begin{bmatrix}
0.978 & 0.019\\
-2.211 & 0.965
\end{bmatrix}
,~B_p = 
\begin{bmatrix}
6.17 \times 10^{-3} \\ 0.617
\end{bmatrix}
,~{C_p}^T =
\begin{bmatrix}
0\\1
\end{bmatrix}
\]

The equivalent DT eigenvalue is $0.97 \pm j 0.21$. The effect of the damping controller is emulated by introducing an additional damping torque component~($\Delta T = 0.06 \,\Delta \omega$). The closed loop DT state-transition matrix for fixed delay of $2h$ is denoted by $A_{C2}$. The non-zero entries of $A_{C2}$ are $A_{C2}$(1:2,1:2)~=~$A_p$, $A_{C2}$(1,4) = $A_{C2}$(1,5) = $-3.71\times 10^{-4}$, $A_{C2}$(2,4) = $A_{C2}$(2,5) = $-3.71\times 10^{-2}$, $A_{C2}$(3,4) = $A_{C2}$(4,5) = $A_{C2}$(5,2)~=~1. $A_{C2}$ has the DT swing-mode eigenvalue pair $ 0.93 \pm j 0.22$. The other three eigenvalues are introduced due to the modelling of the delay. If the minimum and maximum delay are $2h$ and $3h$ respectively, then this forms a switched LTI system with two switching states. The following section presents the stability assessment of switched LTI systems. .



\section{Stability Analysis: Variable Delay Condition}

For a fixed time delay scenario, the system stability can be assessed from the eigenvalues of the closed-loop system matrix $A_c$. Under variable delay conditions, this forms a switched LTI system. Note that the stability of the individual switching state does not guarantee the stability of the switched LTI system~\cite{LMI}. Two methods of assessing the stability of the switched LTI system are presented here. The first approach is based on LMI, which does not depend on the system parameters. Although this is a necessary and sufficient criterion, it cannot directly estimate the damping information of the switched system. In order to overcome this issue, a simplified approximate method is proposed here. Although this approach is based on certain assumptions, they are usually satisfied in practical cases. 

\subsection{LMI Based Approach}
Consider an autonomous switched linear system with $N$ number of linear switching states. The dynamical equation of the $i^{\text{th}}$ switching state is given by $x[K+1] = A_{Ci}\, x[K]$, where $x[K]$ denotes the state variables at the $K^{\text{th}}$ sampling instant.


It is given in~\cite{LMI} that under arbitrary switching conditions, the switched LTI system is stable if there exist $N$ symmetric positive definite matrices $P_1, \cdots ,P_N$, which satisfy
\begin{align} \label{Eq:lmi_approach_switched_lti}
    \begin{bmatrix}
P_i & {A_{Ci}}^T P_j\\
P_j A_{Ci} & P_j
\end{bmatrix}
>0 
\end{align}
for all possible combinations of $(i,j)$. The disadvantages are: (a) the order of the LMI problem becomes very large with a  higher number of switching states, and (b) this method cannot predict the effective damping of the overall switched system. This is particularly important because WADCs are employed to enhance the damping of the inter-area modes. For the SMIB system given in the previous section, the stability of the switched system under the switching states $A_{C2}$ and $A_{C3}$ is  assessed by the LMI based approach. The switched system is stable since there exists a $P$ matrix that  satisfies~\eqref{Eq:lmi_approach_switched_lti}. The eigenvalues of $P$ are $\{0.04,~0.42,~0.81,~1.56,~2.16\}  $, which indicates that $P > 0$. The LMI is solved using the YALMIP solver~\cite{LMI_solver}.

 
\subsection{Proposed simplified stability analysis}

\begin{figure}[!h]
    \centering
    \includegraphics[scale=0.53]{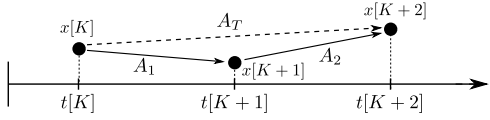}
    \caption{State-transition for switched LTI system}
    \label{Fig:A1_A2}
\end{figure}

Consider the linear switched system with switching states $A_1$ and $A_2$  as shown in Fig. \ref{Fig:A1_A2}. $A_1$ is the state transition matrix from $t[K]$ to $t[K+1]$, while $A_2$ is state transition matrix from $t[K+1]$ to $t[K+2]$.
$A_T = A_2 A_1$ is an effective state matrix from $t[K]$ to $t[K+2]$. In general, if the state transition matrices from $t[0]$ to $t[n]$ are $A_1, A_2,... , A_n$, then the overall state-transition matrix from $t[0]$ to $t[n]$ is given by $A_T = A_n A_{n-1} \cdots A_1$. With the proposed strategy, the switching states have a special structure such that the right and left eigenvectors corresponding to swing mode are approximated to be constant under the following assumptions:\\
(a) $h$ is much smaller than swing mode time period,\\
(b) open-loop swing mode damping ratios are small, \\
(c) modal feedback signal~(as suggested in~\cite{vedanta}) is used.\\
For the sake of brevity, the proof is omitted here. For the SMIB example given in Section II, all the  assumptions are valid. It can be verified that the eigenvector direction corresponding to the swing mode in different switching states is practically constant. The assessment of the stability and estimation of the effective damping of the switched system is presented next.

\par \underline{Stability assessment:} Let the right eigenvector corresponding to the DT swing mode~(of all switching states) be represented by $v$. The following relationship therefore holds.
\begin{align} \label{Eq:switched_matrices_swing_ev_relation}
    A_T \, v \approx (\mu_1 \mu_2 \cdots \mu_n) v = \mu_T v
\end{align}

where $\mu_i$ is the DT swing mode of $A_i$. The overall switched DT system is stable if $\mu_T < 1$, assuming that the other modes~(introduced due to modelling of the delay) are not destabilized by the feedback input. Equation~\eqref{Eq:switched_matrices_swing_ev_relation} implies 
$$|\mu_{min}|^n \leq |\mu_T| \leq |\mu_{max}|^n$$ 
where $\mu_{min}$ and $\mu_{max}$ represent the DT swing modes of the switching states with minimum and maximum absolute values, respectively. Therefore, if all the switching states are individually stable, the switched LTI system will also be stable. 



\par \underline{Damping estimation:} Recall that the absolute value of the DT swing-modes reflects the damping of the CT eigenvalues. The effective damping ($d_e$) of the switched linear system can be defined as the average damping over the entire time duration of $n$ switching states. This can be expressed as given in~\eqref{Eq:damping_switched_systems}.
\begin{align} 
d_e &= |\mu_T|^{\frac{1}{n}} = \Big|\mu_T^{\frac{1}{n}}\Big| \implies |\mu_{min}| \leq d_e \leq  |\mu_{max}|  \label{Eq:damping_switched_systems}
\end{align}
 From~\eqref{Eq:damping_switched_systems}, it can be seen that the effective damping of the switched system is bounded by the minimum and maximum damping among individual switching states. 


\section{Case Study: 2-area 4-machine System}
\begin{figure}[!h]
    \centering
    \includegraphics[scale=0.34]{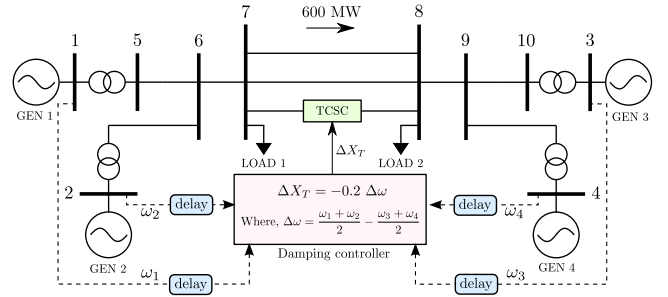}
    \caption{Schematic of 4-machine system with damping controller}
    \label{Fig:SMIB}
\end{figure}

Consider the case of the 2-area 4-machine system connected via three parallel transmission lines. The generator and network parameters are taken from~\cite{kundur}. The generators are equipped with a static exciter~(without TGR) model given in~\cite{kundur}. One of the parallel lines is equipped with a Thyristor controlled series capacitor (TCSC), which is modelled as a variable reactance. The TCSC modulates its reactance in proportion to the difference between the average frequency  of the two areas~($\Delta \omega$), as shown in Fig.~\ref{Fig:SMIB}.  Note that the negative gain of the controller reflects the inverse correlation between the reactance and power flow through the line. The base frequency of the system is 60~Hz. The PMU sampling rate~($h$) is 16.67~ms, corresponding to one sample per cycle. Note that the discretization time-step is also equal to $h$.

\par
\begin{figure}[!h]
    \centering
    \includegraphics[scale=0.175]{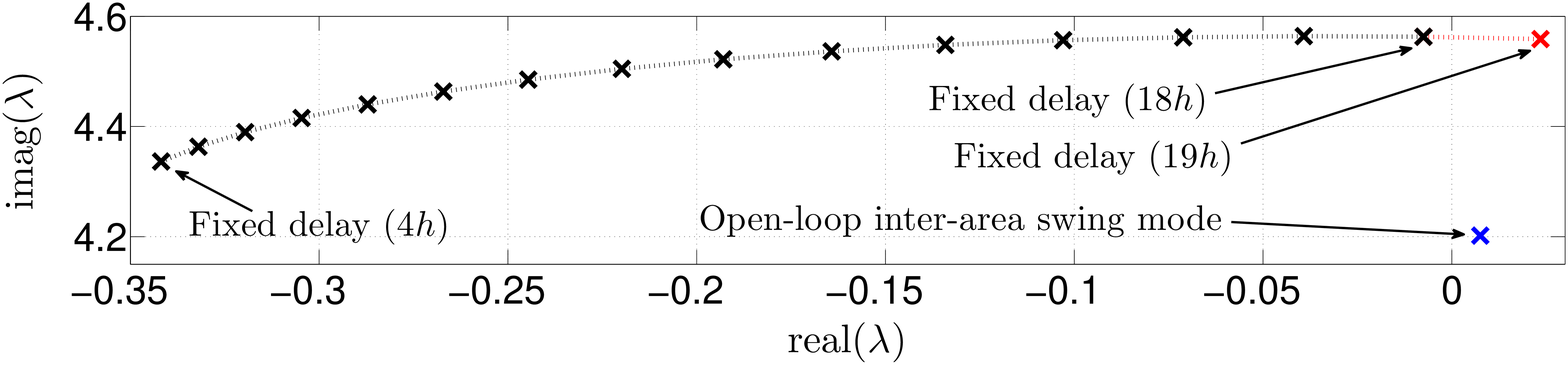}
    \caption{Root-locus of CT swing mode}
    \label{Fig:eig_trac}
\end{figure}

 The open-loop system~(without damping controller) has an unstable CT inter-area swing mode at  $\lambda = 0.007 \pm j4.2$. The objective is to improve the damping of this unstable mode by using wide-area frequency measurements as inputs to the damping controller. Consider that the minimum delay is 66.67~ms~(4$h$) and the maximum permissible delay is 300~ms~(18$h$) i.e. a total of 15 switching states are possible~(excluding the open-loop system). Fig.~\ref{Fig:eig_trac} shows root-locus of the CT inter-area swing mode eigenvalue~[obtained from the corresponding DT eigenvalue using~\eqref{Eq:trapezoidal_rule}] for multiple fixed delays~(corresponding to individual switching states). Note that all the individual switching states are stable. The maximum permissible delay is chosen to be $18h$ as the system is unstable with a fixed delay of 19$h$.



\par \underline{LMI Based Approach:} The stability of switched system~(with the 15 switching states) is first assessed by the LMI based method. \textcolor{blue}{It has been found that the switched system satisfies the LMI condition given in~\eqref{Eq:lmi_approach_switched_lti}, with a single $P$ matrix. Therefore, the switched system is stable under variable delay between 4$h$ and 18$h$.} 
\par
\begin{figure}[!h]
    \centering
    \includegraphics[scale=0.28]{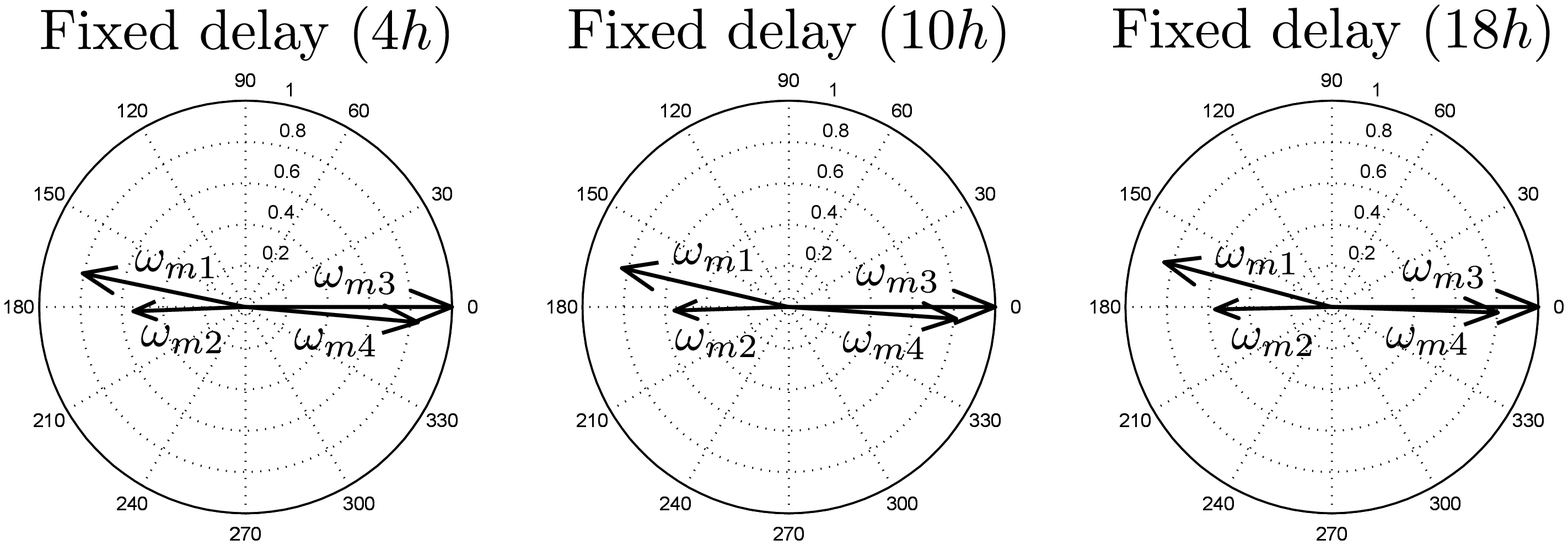}
    \caption{Inter-area swing mode eigenvector: generator speed components}
    \label{Fig:eig_vec}
    \vspace*{3mm}
    \includegraphics[scale=0.175]{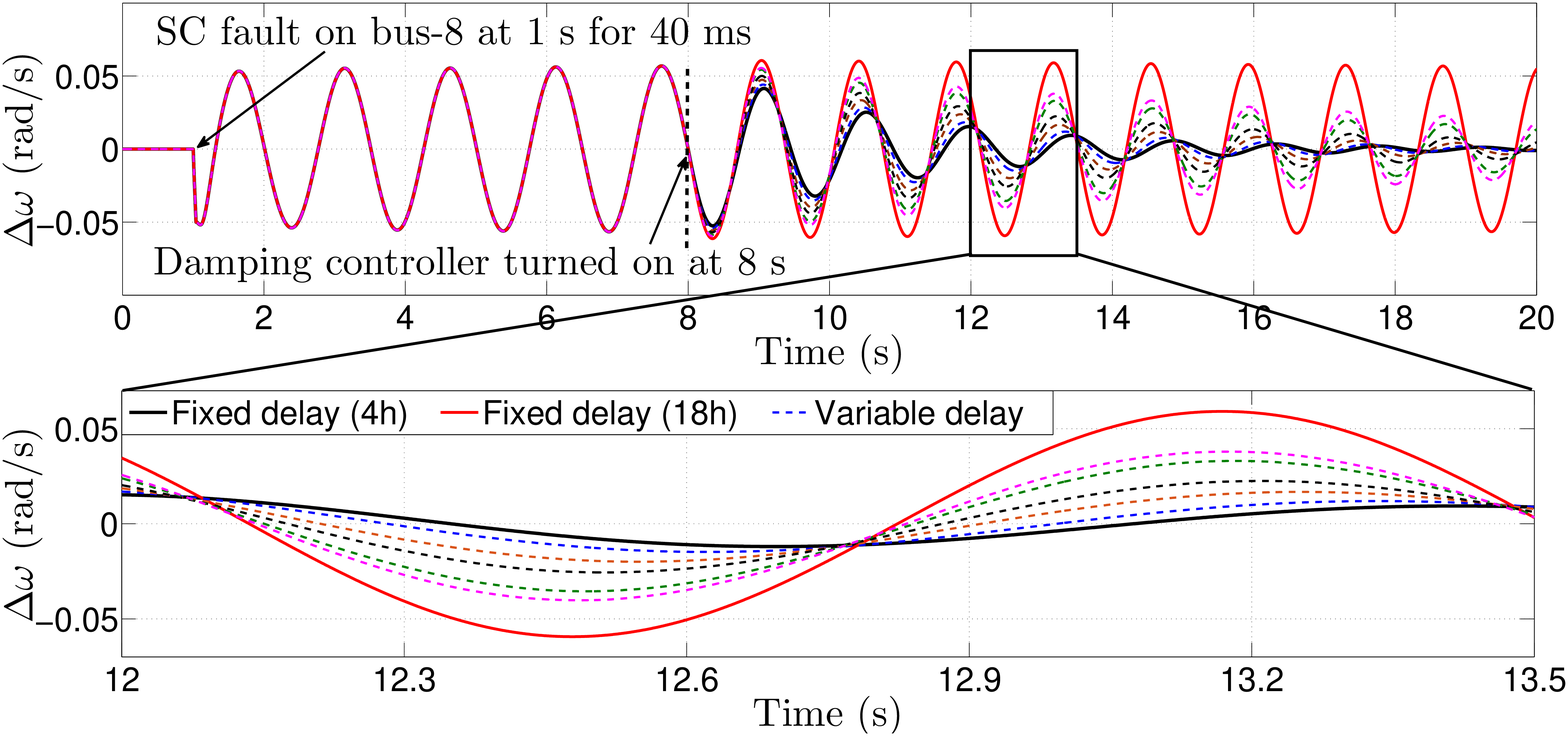}
    \includegraphics[scale=0.175]{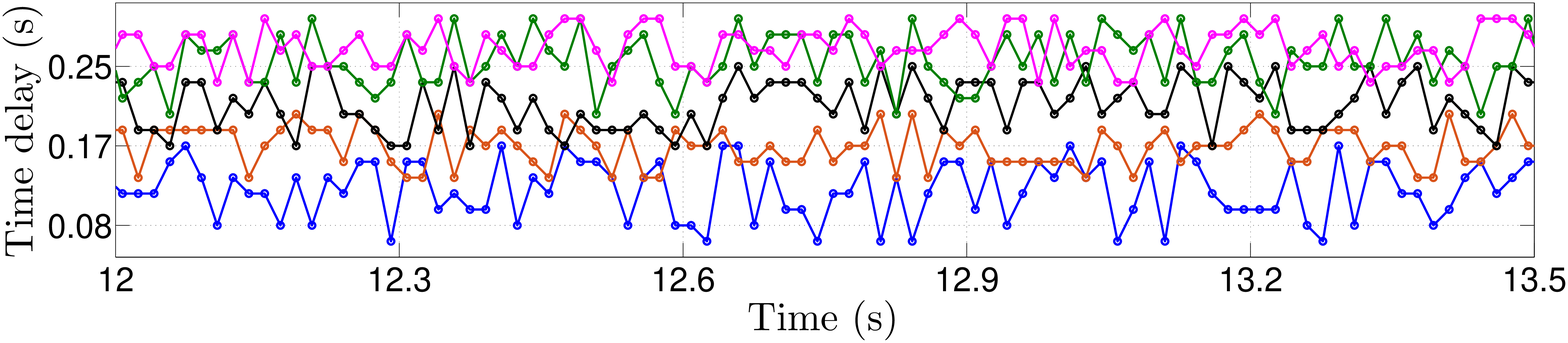}
    \caption{Time-domain simulations of the system with variable delay}
    \label{Fig:dw_plot1}
\end{figure}

 \underline{Proposed Approach:} The system considered here satisfies all the assumptions stated in Section~III. In order to assess the applicability of the proposed approach, the components of the right eigenvector of the inter-area swing mode corresponding to the generator speed state variables, under different fixed delay conditions, are plotted in Fig.~\ref{Fig:eig_vec}. It can be seen that the relative mode-shape of the inter-area swing mode can be considered to be constant under different delay conditions. 
\par Since all the individual switching states corresponding to different fixed delays are individually stable, the overall system will also be stable, if the delay is allowed to vary between the specified limits~(4$h$ and 18$h$).  The root loci in Fig.~\ref{Fig:eig_trac} suggests that the damping is minimum~($0.16\%$) for 18$h$ fixed delay, and is maximum~($7.86\%$) when the fixed delay is 4$h$. Therefore, the effective damping of the system under the variable delay situation should be bounded between these limits. 
\par To validate the proposed analysis, time-domain simulations are performed on the given system. The bus frequency measurements are delayed with random delays between 4$h$ and 18$h$. The simulation is carried out with several realizations of the random delay process, as shown in Fig.~\ref{Fig:dw_plot1}. To check the stability of the overall  system, a symmetrical short-circuit fault at bus~8~(of duration 40~ms) is applied at $t = 1$~s. The TCSC damping controller is initially kept disabled until $t= 8$~s. The variation in $\Delta \omega$ is shown in Fig.~\ref{Fig:dw_plot1} for the given delay realizations. It can be seen that the unstable inter-area mode of the open-loop system starts growing till the TCSC damping controller is enabled. The response of the system for fixed delays corresponding to the minimum and maximum damping is also given here. \textcolor{blue}{It can be seen that the system under the specified variable delay conditions is stable, as predicted by the proposed approach. Furthermore, the effective damping ratio~(for the different delay realizations given here) are found to be $\{2.72\%, 3.82\%,4.27\%,5.53\%,6.42\%\}$,  which are bounded by the minimum and maximum damping provided by the individual switching states.}

\section{Conclusion and Future work}

This paper presents a data-handling strategy under variable time-delay conditions in wide-area feedback and control. \textcolor{blue}{The advantages of the proposed strategy are that (i) the stability of the closed-loop system under variable delay conditions can be easily predicted, and (ii) the bounds on the effective damping of the variable-delay system can be estimated from that of the individual switching states.} The simplified analysis gives accurate results if (a) the discretization time-step is much smaller than the swing mode time period, (b) open-loop swing mode damping ratios are small, and (c) a modal feedback signal is used. These are verified using a case study, which attests to the effectiveness of the proposed analysis. Further work is required to evaluate the proposed strategy for large realistic systems and with multi-modal feedback signals and multiple actuators.



\medskip
\bibliographystyle{IEEEtran}
\bibliography{ref}

\begin{thebibliography}{10}
\providecommand{\url}[1]{#1}
\csname url@samestyle\endcsname
\providecommand{\newblock}{\relax}
\providecommand{\bibinfo}[2]{#2}
\providecommand{\BIBentrySTDinterwordspacing}{\spaceskip=0pt\relax}
\providecommand{\BIBentryALTinterwordstretchfactor}{4}
\providecommand{\BIBentryALTinterwordspacing}{\spaceskip=\fontdimen2\font plus
\BIBentryALTinterwordstretchfactor\fontdimen3\font minus
  \fontdimen4\font\relax}
\providecommand{\BIBforeignlanguage}[2]{{%
\expandafter\ifx\csname l@#1\endcsname\relax
\typeout{** WARNING: IEEEtran.bst: No hyphenation pattern has been}%
\typeout{** loaded for the language `#1'. Using the pattern for}%
\typeout{** the default language instead.}%
\else
\language=\csname l@#1\endcsname
\fi
#2}}
\providecommand{\BIBdecl}{\relax}
\BIBdecl

\bibitem{vedanta}
V.~Pradhan, A.~M. Kulkarni, and S.~A. Khaparde, ``{A Composite Strategy for
  Power Oscillation Damping Control Using Local and Wide Area Feedback
  Signals},'' \emph{IEEE Trans. Power Syst.}, vol.~31, no.~3, pp. 2348--2360,
  2016.

\bibitem{delay_stab}
L.~Cheng \emph{et~al.}, ``{Adaptive Time Delay Compensator (ATDC) Design for
  Wide-Area Power System Stabilizer},'' \emph{IEEE Trans. Smart Grid}, vol.~5,
  no.~6, pp. 2957--2966, 2014.

\bibitem{d_2015}
F.~Zhang \emph{et~al.}, ``{Measurement and Modeling of Delays in Wide-Area
  Closed-Loop Control Systems},'' \emph{IEEE Trans. Power Syst.}, vol.~30,
  no.~5, pp. 2426--2433, 2015.

\bibitem{fd2_2008}
Z.~Liu, C.~Zhu, and Q.~Jiang, ``{Stability analysis of time delayed power
  system based on Cluster Treatment of Characteristic Roots method},'' in
  \emph{2008 IEEE Power Energy Soc. Gen. Meet.}, 2008, pp. 1--6.

\bibitem{wilches2017effect}
F.~Wilches-Bernal \emph{et~al.}, ``{Effect of time delay asymmetries in power
  system damping control},'' in \emph{2017 IEEE Power Energy Soc. Gen.
  Meet.}\hskip 1em plus 0.5em minus 0.4em\relax IEEE, 2017, pp. 1--5.

\bibitem{d_2014}
W.~Yao \emph{et~al.}, ``{Wide-Area Damping Controller of FACTS Devices for
  Inter-Area Oscillations Considering Communication Time Delays},'' \emph{IEEE
  Trans. Power Syst.}, vol.~29, no.~1, pp. 318--329, 2014.

\bibitem{delay_em1}
C.~F.~M. Danielson \emph{et~al.}, ``{Analysis of communication network
  challenges for synchrophasor-based wide-area applications},'' in \emph{2013
  IREP Symposium Bulk Power System Dynamics and Control - IX Optimization,
  Security and Control of the Emerging Power Grid}, 2013, pp. 1--13.

\bibitem{data_dropout}
J.~Wu and T.~Chen, ``{Design of Networked Control Systems With Packet
  Dropouts},'' \emph{IEEE Trans. Automat. Contr.}, vol.~52, no.~7, pp.
  1314--1319, 2007.

\bibitem{data_disordering}
Y.-B. Zhao, G.-P. Liu, and D.~Rees, ``{Design of a Packet-Based Control
  Framework for Networked Control Systems},'' \emph{IEEE Trans Control Syst.
  Technol.}, vol.~17, no.~4, pp. 859--865, 2009.

\bibitem{kevin}
K.~K. Gajjar and A.~M. Kulkarni, ``{Selective Eigenvalue Analysis for Wide-Area
  Measurement and Control Applications – A Review},'' in \emph{2019 8th
  International Conference on Power Systems (ICPS)}, 2019, pp. 1--6.

\bibitem{LMI}
J.~Daafouz, P.~Riedinger, and C.~Iung, ``{Stability analysis and control
  synthesis for switched systems: a switched Lyapunov function approach},''
  \emph{IEEE Trans. Automat. Contr.}, vol.~47, no.~11, pp. 1883--1887, 2002.

\bibitem{LMI_solver}
J.~L{\"{o}}fberg, ``{YALMIP : A Toolbox for Modeling and Optimization in
  MATLAB},'' in \emph{CACSD Conference}, Taipei, Taiwan, 2004.

\bibitem{kundur}
M.~Klein, G.~Rogers, and P.~Kundur, ``{A fundamental study of inter-area
  oscillations in power systems},'' \emph{IEEE Trans. Power Syst.}, vol.~6,
  no.~3, pp. 914--921, 1991.

\end{thebibliography}

\end{document}